\documentstyle[preprint,aps,epsf]{revtex}
\tighten
\begin{document}
\draft

\title{Testing nuclear forces by polarization transfer coefficients in 
$d(\overrightarrow p, \overrightarrow p)d$ and 
$d(\overrightarrow p, \overrightarrow d)p$ reactions at $E^{lab}_p = 22.7$~MeV
}

\author{
H.~Wita\l{}a$^1$, J.Golak$^1$, R.~Skibi\'nski$^1$, W.~Gl\"ockle$^2$,
A.~Nogga$^3$, E. Epelbaum$^4$, H.~Kamada$^5$, A. Kievsky$^6$, M. Viviani$^6$}

\address{$^1$M. Smoluchowski Institute of Physics, Jagiellonian University,
                    PL-30059 Krak\'ow, Poland}
\address{$^2$Institut f\"ur Theoretische Physik II,
         Ruhr-Universit\"at Bochum, D-44780 Bochum, Germany}
\address{$^3$ Institut f\"ur Kernphysik, Forschungszentrum J\"ulich,
D-52425 J\"ulich, Germany}
\address{$^4$ Jefferson Laboratory, Theory Division, Newport News,
VA 23606, USA}

\address{$^5$ Department of Physics, Faculty of Engineering,
   Kyushu Institute of Technology,
   1-1 Sensuicho, Tobata, Kitakyushu 804-8550, Japan}
\address{$^6$ Istituto Nazionale di Fisica Nucleare, Via Buonarroti 2,
I-56100 Pisa, Italy}

\date{\today}

\maketitle

\begin{abstract}
The proton to proton polarization transfer coefficients 
$K_x^{x'}$, $K_y^{y'}$, 
$K_z^{x'}$ and the proton to deuteron polarization transfer coefficients 
$K_x^{x'}$, $K_y^{y'}$, 
$K_z^{x'}$, $K_x^{y'z'}$, $K_y^{z'z'}$, 
$K_z^{y'z'}$, $K_y^{x'z'}$ and $K_y^{x'x'-y'y'}$ have been measured in 
 $d(\overrightarrow p, \overrightarrow p)d$ and 
$d(\overrightarrow p, \overrightarrow d)p$ reactions at $E^{lab}_p = 22.7$~MeV,
 respectively. The data have been compared to predictions of modern nuclear 
forces obtained by solving the three-nucleon Faddeev equations in 
momentum space.  Realistic (semi) phenomenological nucleon-nucleon 
potentials combined with model three-nucleon forces and modern 
chiral nuclear forces have been used. The AV18, CD Bonn, Nijm I and II 
nucleon-nucleon interactions have been applied alone or combined 
with the Tucson-Melbourne 99 three-nucleon force, adjusted separately 
for each potential to reproduce the triton binding energy. For the AV18 
potential also the Urbana IX three-nucleon force have been used. In addition 
chiral NN 
potentials in the next-to-leading-order and chiral two- and three-nucleon forces 
in the next-to-next-to-leading-order have been applied. 
Only when three-nucleon forces are included a satisfactory description of all data 
results. For the chiral approach 
 the restriction to the forces in the next-to-leading 
order is insufficient. Only when going over to the next-to-next-to-leading 
order one gets a satisfactory description of the data, similar to the one obtained with 
the (semi) phenomenological forces. 

\end{abstract}


\section{Introduction}

A rich set of observables provided by three-nucleon (3N) 
reactions can be used to test 
modern nuclear forces. Presently there are two theoretical approaches to 
construct them. In the traditional approach the nucleon-nucleon (NN) 
potentials are derived in the framework of the meson-exchange picture alone or 
mixed with   phenomenological assumptions. 
By adjusting parameters so called realistic, 
high precision  interactions such as the phenomenological AV18~\cite{AV18} 
potential  
  and the meson-theoretical CD Bonn~\cite{CDBONN} 
 together with  Nijm I and  II ~\cite{NIJMI}  
potentials are obtained. They  provide a very good description of NN data 
below about $350$~MeV nucleon laboratory energy. All these potentials 
have in common that they fit the large set of NN data with $\chi^2$ per 
datum close to one, indicating essentially phase equivalence.

In a more modern framework of chiral effective field theory, nuclear
forces are
linked to the underlying strong interaction between quarks and gluons.
They are derived from the most general effective Lagrangian for pions and
nucleons, which is consistent with the spontaneously broken approximate
chiral symmetry of QCD, using chiral perturbation theory 
($\chi$PT)~\cite{wein91}. 
The $\chi$PT approach gives a deeper understanding of nuclear forces
than the traditional approach and allows to construct three-- and
more--nucleon
forces consistent with the NN interactions \cite{vankolck94,epel98,epel03}.
In practice, various contributions to the nuclear force in the $\chi$PT
framework
are organized in terms of the expansion in $Q/\Lambda$,
where $Q$ is the soft scale corresponding to the nucleon external
momenta and the
pion mass and $\Lambda$ is the hard scale associated with the chiral
symmetry breaking
scale or an ultraviolet cut--off. At present, the two--nucleon system
has been studied
up to next--to--next--to--next--to--leading order (NNNLO) in the chiral
expansion \cite{epel03,nnnlo}.
At this order, the two--nucleon force receives the contributions from
one--pion exchange,
two--pion exchange at the two--loop level as well as three--pion
exchange, which turn
out to be numerically irrelevant. In addition, one has to take into
account all possible
short--range contact interactions with up to four derivatives and the
appropriate
isospin--breaking effects. In the three-- and more--nucleon sectors, the
calculations have
so far only been performed up to next--to--next--to--leading order
(NNLO) \cite{epel03}. In this work, we
will show the results corresponding to the latest version of the chiral
NN forces introduced
in references \cite{epel04a,nnnlo} and based on the spectral function
regularization scheme~\cite{epel04b}.

Recent studies of few-nucleon bound states and of 3N reactions provided 
numerous indications that three-nucleon forces (3NFs) form an important 
component of the potential energy of three interacting 
nucleons~\cite{Friar1993,Nogga1997,Argonne,wit98,glo96,nogga03}. In the 
traditional approach they are accounted for by adding 
model 3NFs, such as e.g. the $2\pi$-exchange 
Tucson-Melbourne (TM)~\cite{TM} or Urbana IX~\cite{uIX} interactions, 
with parameters 
adjusted to reproduce the experimental 
triton binding energy. Such a simple treatment allows to cure some of the 
discrepancies between data and 
theory~\cite{sek02,wit01,abf98,wit99,hat02,cad01,KVR01}. 
In the approach based on $\chi$PT nonvanishing 3NFs appear in the 
next-to-next-to-leading order (NNLO) of the chiral expansion. 
In addition to the $2\pi$-exchange term 
two other  topologies appear. One of them corresponds to a contact 
interaction of three nucleons and the second  to a contact interaction 
of two nucleons exchanging in addition one pion with the third nucleon. 
The two free parameters of these two terms are adjusted by fitting 
two independent 3N observables 
(e.g. the triton binding and the nd doublet scattering length). 
The quality of the description of the 3N observables is then similar in both 
approaches~\cite{epel2002}. 
 The study of the details of the 3NF's is a lively 
topic  of present day few-nucleon system studies. 

In the present paper we would like to analyze the proton to proton 
and the proton to deuteron spin transfer coefficients measured in 
$d(\overrightarrow p, \overrightarrow p)d$ and 
$d(\overrightarrow p, \overrightarrow d)p$ reactions, respectively, 
 at $E^{lab}_p = 22.7$~MeV~\cite{glomb,kretch,kretch1}. 
The existing data for polarization transfer coefficients 
in elastic nucleon-deuteron (Nd) scattering are restricted to 
few experiments in the pd system ($E_p^{lab}=10$~MeV~\cite{sperisen}, 
$E_p^{lab}=19$~MeV~\cite{sydow}, 
$E_p^{lab}=250$~MeV~\cite{hat02}, $E_d^{lab}=270$~MeV~\cite{sek04}) and to 
one measurement in the nd system ($E_p^{lab}=19$~MeV~\cite{vwitsch}).  
After a  short description of the  theoretical calculations in 
Section II in section III we show the data  and compare them to different 
theoretical predictions. The summary and conclusions follow in Section IV.


\section{Calculations}

In this work we will employ two different methods to solve the 
nucleon-deuteron scattering problem. Our first scheme is based on Faddeev 
equations. 
The nucleon-deuteron  elastic scattering with neutron and protons interacting
through a NN potential $V$ and through a 3NF $V_4$ 
is described in terms of a breakup operator T
satisfying the Faddeev-type integral equation~\cite{wit88,glo96,hub97}

\begin{eqnarray}
T &=& t P  + (1 + tG_0)V_4^{(1)}(1 + P) +  t P G_0 T  +  
(1 + tG_0)V_4^{(1)}(1 + P)G_0T .
\label{eq1a}
\end{eqnarray}
The two-nucleon (2N) t-matrix t results from the interaction $V$ through
the Lippmann-Schwinger equation. 
The permutation operator  $P=P_{12}P_{23} + P_{13}P_{23}$ is given in terms
of the transposition $P_{ij}$ which interchanges nucleons i and j 
 and $G_0$ is the free 3N propagator.  
 Finally the operator $V_4^{(1)}$ appearing in Eq.(\ref{eq1a}) is part of the 
full 3NF $V_4 = V_4^{(1)} + V_4^{(2)} + V_4^{(3)}$ and is symmetric under 
exchange of nucleons 2 and 3. For instance, in the case of the $\pi-\pi$ 
exchange 3NF such a decomposition corresponds to the three possible 
choices of the nucleon which undergoes off-shell $\pi-N$ scattering. 
It is understood that the operator T acts on 
 the incoming state  
$ \vert \phi > =
\vert \vec q_0 > \vert \phi_d > $ which describes the free
nucleon-deuteron motion with relative momentum $\vec q_0$ and
the deuteron wave function $\vert \phi_d >$.  
 The physical picture underlying Eq.(\ref{eq1a})  is
revealed after iteration which  leads to  a multiple scattering series 
 for T.  

The  elastic Nd scattering transition operator U 
 is  given in terms of T by~\cite{wit88,glo96,hub97}
\begin{eqnarray}
U  &=& P G_0^{-1} + P T + V_4^{(1)}(1 + P) + V_4^{(1)}(1 + P)G_0T .
\label{eq1c}
\end{eqnarray}

We solve Eq.(\ref{eq1a}) in momentum space using a partial wave decomposition 
for each total angular momentum $J$ and parity of the 3N system. To achieve 
converged results a sufficiently high number of partial waves have been 
used. Calculations with and without 3NF were performed including all 3N 
partial wave states with total two-body angular momenta up to $j=5$. In the case 
when the 3NF is switched-off Eq.(\ref{eq1a}) is solved for $J$ up to $25/2$. 
When the shorter ranged 3NF is also active it is sufficient to go up to 
$J \le 13/2$ only. In all calculations we neglect the total isospin $T=3/2$ 
contribution in the $^1S_0$ state and use in 
this state a np form of the NN interaction. 
Such a restriction to the np force for the $^1S_0$ state does not have 
a significant effect on the polarization transfer coefficients. 

The second scheme is  based on the Kohn Variational principle, the
$S$--matrix elements corresponding to a 3N scattering state with total angular
momentum $J$ can be obtained as the stationary point of the functional
\begin{equation}
[{}^J{S}^{SS'}_{LL'}]= {}^J{S}^{SS'}_{LL'}+{i}
\langle\Psi^-_{LSJ}|H-E|\Psi^+_{L'S'J}\rangle \ .
\label{eq:ckohn}
\end{equation}
The wave function $\Psi^+_{LSJ}$ describes a 3N scattering state in which
asymptotically an ingoing nucleon is approaching the deuteron in a
relative angular momentum $L$ and total spin $S$. The parity of the state
is given by $(-1)^L$. The wave function is expanded, using a partial
wave decomposition, in terms of the pair
correlated hyperspherical harmonic (PHH) basis as described in
Ref.~\cite{KVR01}. As in the Faddeev scheme, states up to $J=25/2$
have been considered.

\begin{figure}
\epsfysize=180mm \epsffile{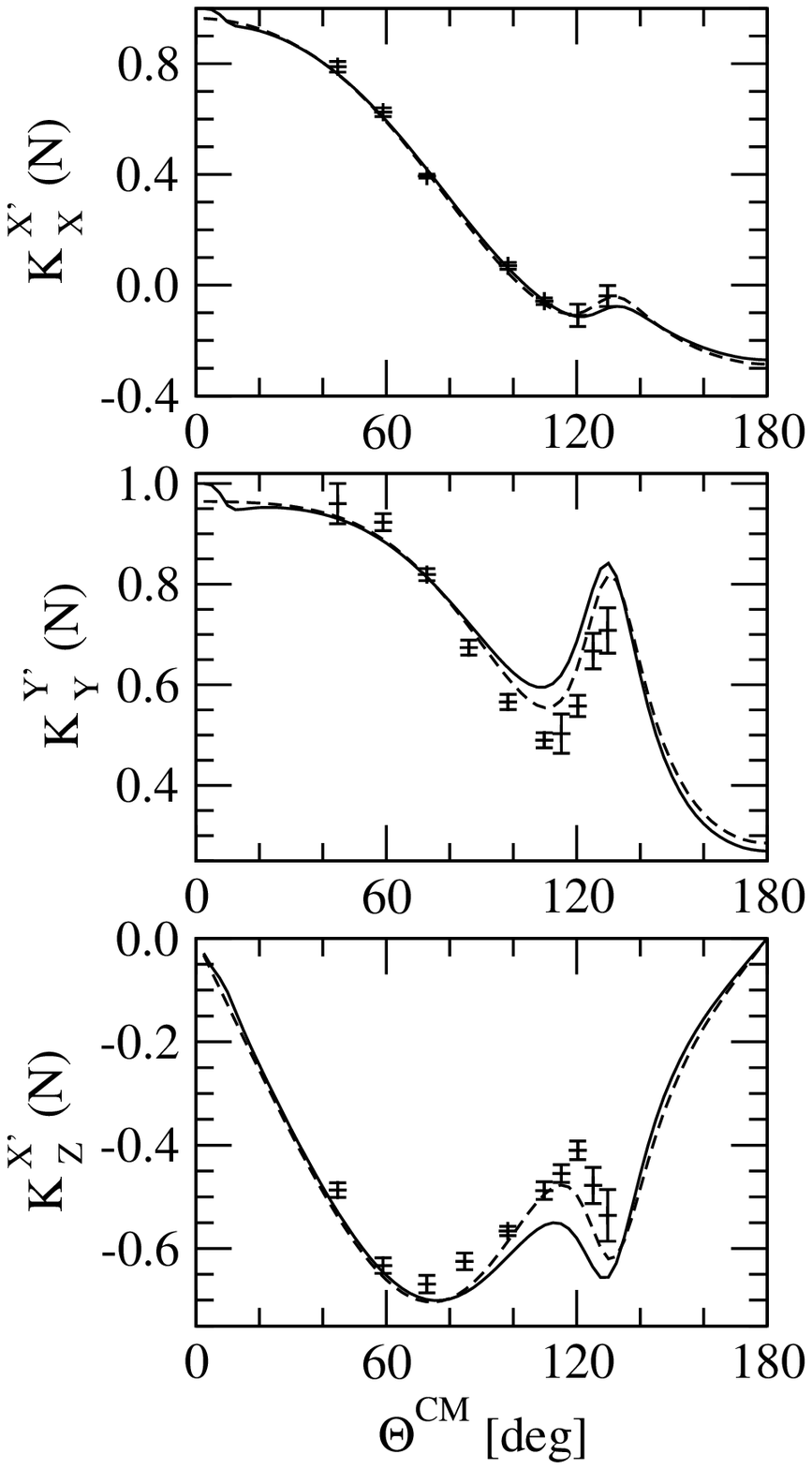}
\caption[]{
The nucleon to nucleon spin transfer coefficients in Nd elastic scattering at 
$E_{lab}^N = 22.7$~MeV. The crosses are the pd data 
 from~\cite{glomb,kretch,kretch1}. 
The dashed line is the result of the hyperspherical harmonic expansion method 
with AV18 potential. 
The solid line is the corresponding result when the  pp Coulomb force 
is included.
}
\label{fig0a}
\end{figure}

\begin{figure}
\epsfysize=180mm \epsffile{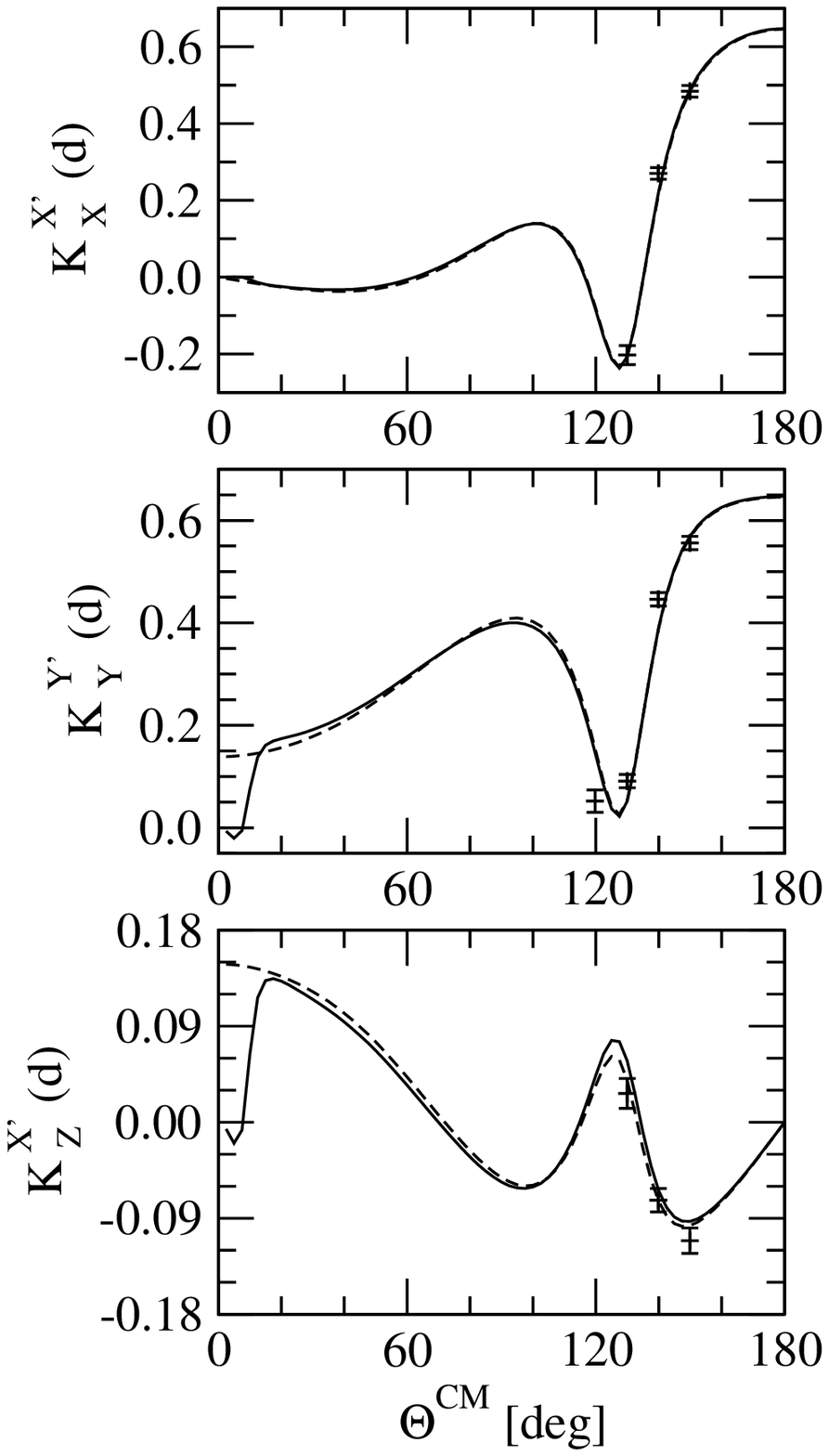}
\caption{
The nucleon to deuteron spin transfer coefficients in Nd elastic scattering at 
$E_{lab}^N = 22.7$~MeV. The description of symbols and lines 
 is the same as 
 in Fig.\ref{fig0a}.
}
\label{fig0b}
\end{figure}

\begin{figure}
\epsfysize=150mm \epsffile{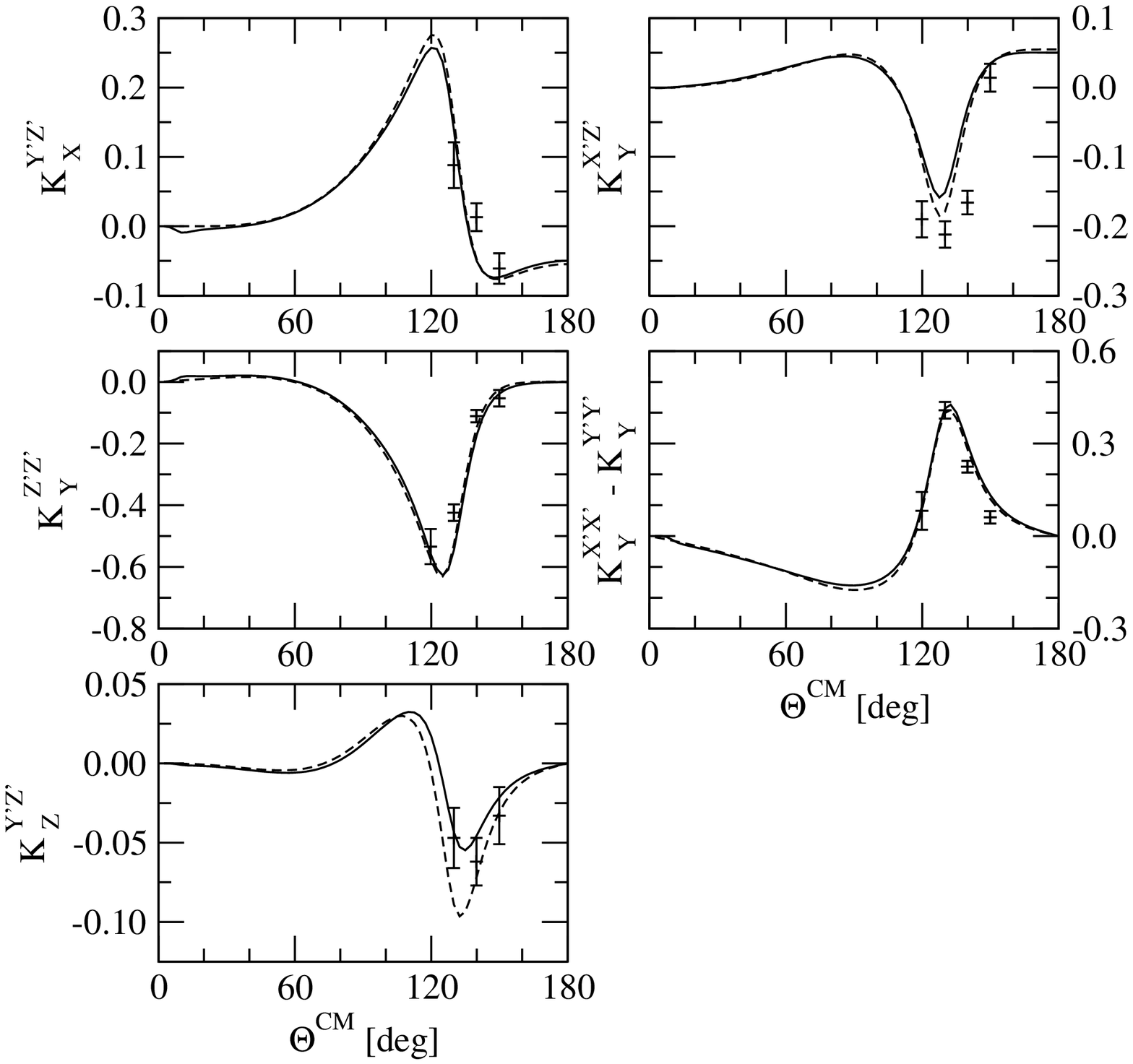}
\caption{
The nucleon  to deuteron spin transfer coefficients in Nd elastic 
scattering at 
$E_{lab}^N = 22.7$~MeV. The description of symbols and lines 
 is the same as 
 in Fig.\ref{fig0a}.
}
\label{fig0c}
\end{figure}

In our Faddeev calculations the Coulomb interaction between two protons 
is totally 
neglected. A measurement of the neutron to neutron polarization transfer 
coefficients $K_y^{y'}$ in neutron-deuteron (nd) elastic 
scattering~\cite{vwitsch}   and their comparison 
to the corresponding pd data~\cite{sydow} shows  
that effects caused by the Coulomb force for this coefficient are 
non-negligible. These effects have been studied on a few polarization transfer
coefficients~\cite{kievsky01} as well as in other polarization observables
\cite{KVR01} using the Kohn variational principle in conjunction with the 
PHH basis. In these calculations the Coulomb force between the two protons
has been considered without approximations and the results confirm sizable
Coulomb-force effects in the energy range considered here.
Therefore to remove  possible  Coulomb-force effects for 
the studied polarization transfer coefficients 
we proceed in the following manner. Using the 
 PHH expansion we evaluate the studied polarization 
transfer coefficients without and with Coulomb force and employ the AV18 
NN interaction. This is displayed in Figs.~\ref{fig0a}-\ref{fig0c}. 
In this manner we read off 
the shifts caused by the pp Coulomb force. Then we generate ``nd'' data by 
applying those shifts to our pd data. 
 For the studied polarization transfers the Coulomb force effects 
are restricted mostly to forward angles and to the region around 
$\theta_{cm} \approx 120^o$. For the proton to proton spin transfer 
coefficient $K_x^{x'}$ they are of minor importance whereas for $K_y^{y'}$ 
and $K_z^{x'}$ the Coulomb force effects  change significantly  the magnitude 
of these coefficients around $\theta_{cm} \approx 120^o$ 
(see Fig.~\ref{fig0a}). For the proton to vector-deuteron spin transfers 
Coulomb force effects are rather small with exception of very forward 
angles, where  they decrease significantly 
 the magnitude  (see Fig.~\ref{fig0b}). For $K_z^{x'}$ 
also some effects are seen around $\theta_{cm} \approx 120^o$. In case 
of the proton to tensor-deuteron transfers 
 shown in Fig.~\ref{fig0c} only $K_x^{y'z'}$ (in the steep slope), 
$K_y^{x'z'}$ and 
 $K_z^{y'z'}$ exhibit large Coulomb force effects in the region of cm angles 
around $\theta_{cm} \approx 120^o$. 
 In the following figures we include both, the  pd and ``nd'' data.

\section{Results}

\begin{figure}
\epsfysize=180mm \epsffile{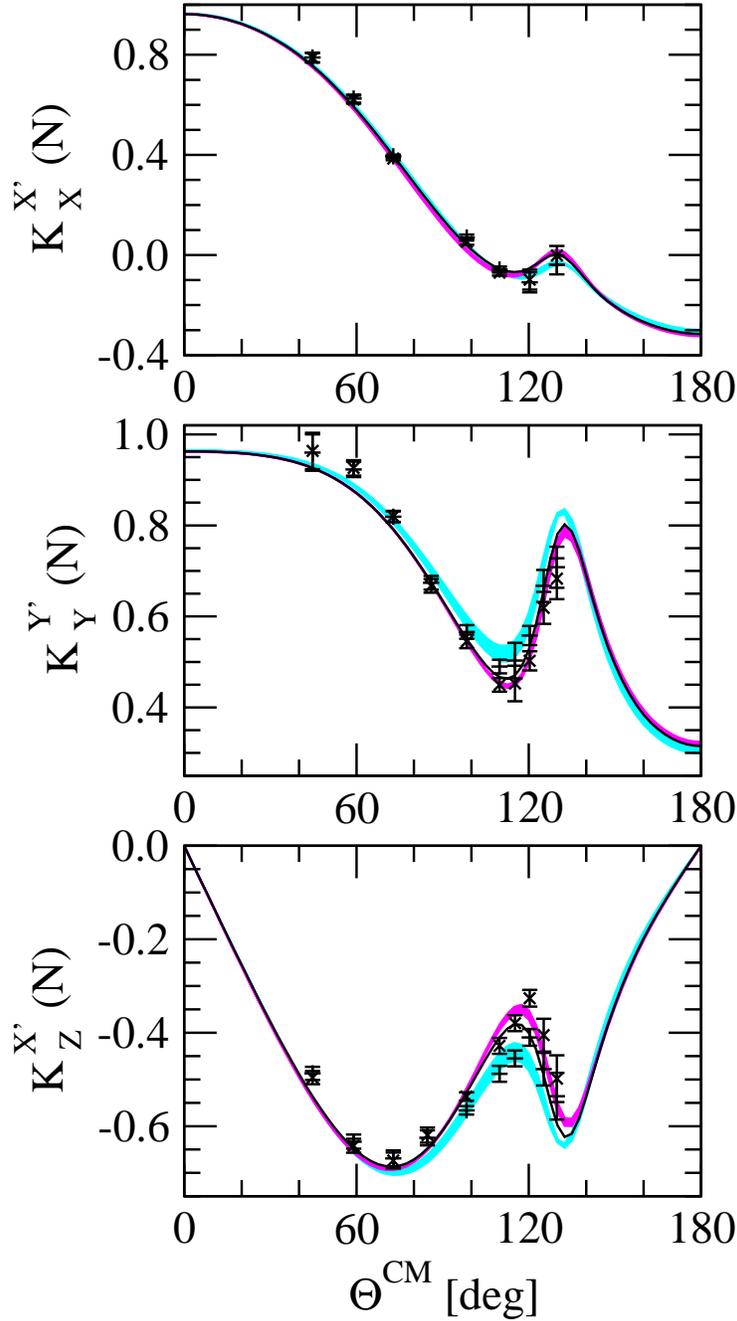}
\caption[]{
The nucleon to nucleon spin transfer coefficients in Nd elastic scattering at 
$E_{lab}^N = 22.7$~MeV. The crosses are the pd data 
from~\cite{glomb,kretch,kretch1} and x-es are 
the corresponding 
``nd data'' obtained by subtracting the effects due to 
the pp Coulomb force (see text for explanation). 
The light band results from theoretical predictions obtained with 
the  AV18, CD Bonn, Nijm I and II NN 
potentials. The dark band is obtained when these interactions 
are combined with 
the TM99 3NF. The solid line is the prediction of the AV18 + Urbana IX 3NF. 
}
\label{fig1}
\end{figure}

\begin{figure}
\epsfysize=180mm \epsffile{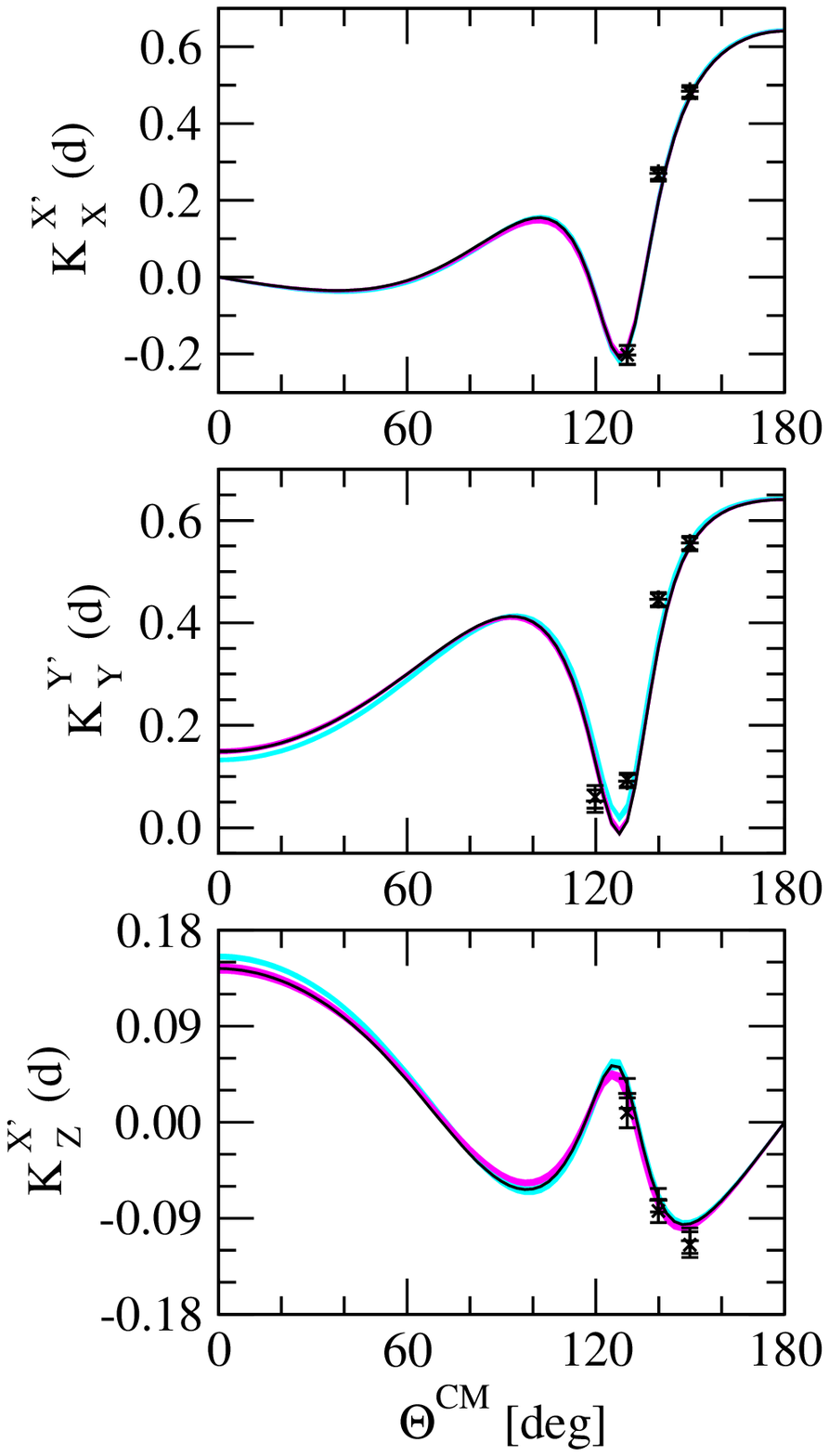}
\caption{
The nucleon to deuteron spin transfer coefficients in Nd elastic scattering at 
$E_{lab}^N = 22.7$~MeV. The description of symbols, bands and lines 
 is the same as 
 in Fig.\ref{fig1}.
}
\label{fig2}
\end{figure}

\begin{figure}
\epsfysize=150mm \epsffile{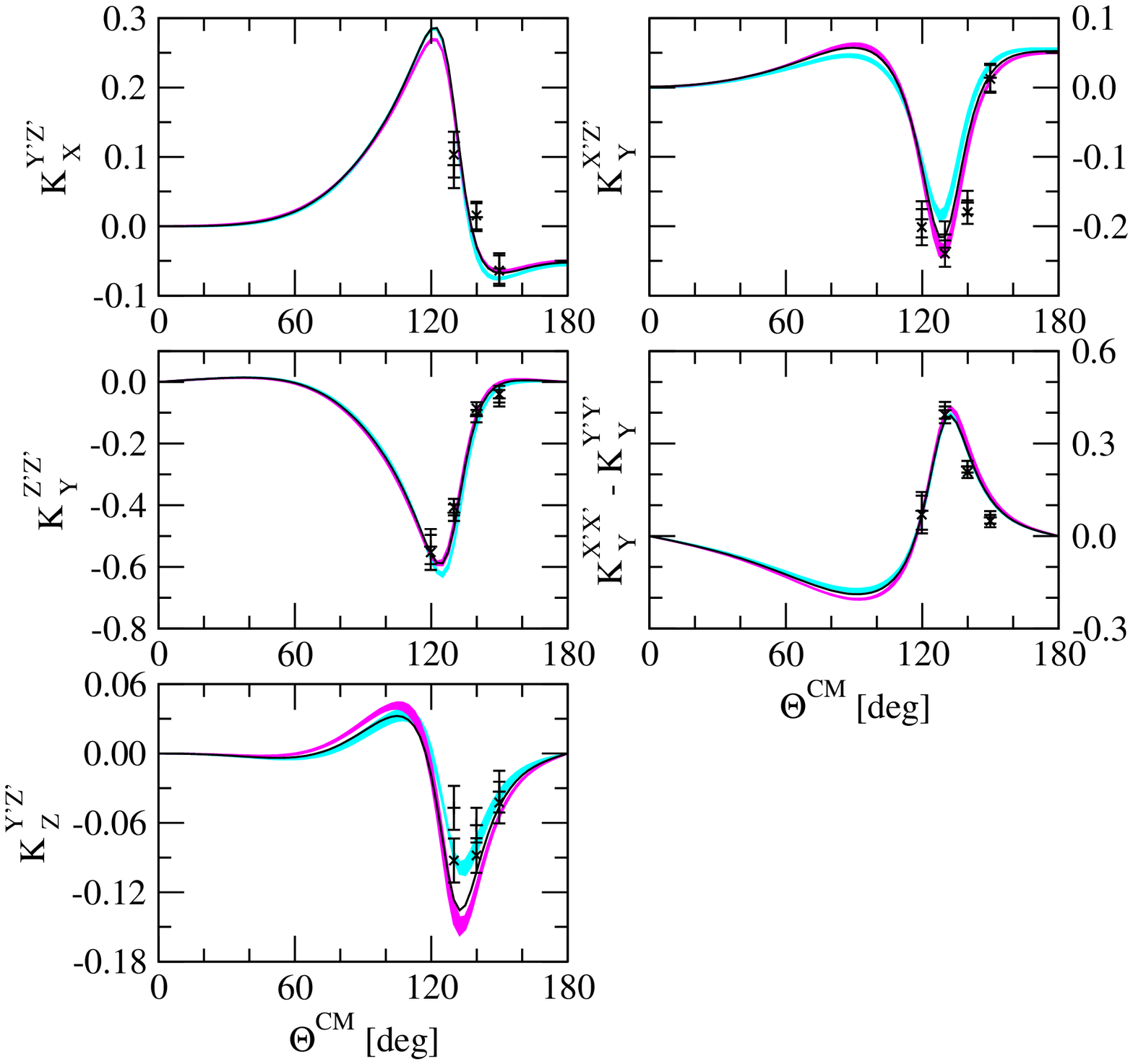}
\caption{
The nucleon to deuteron spin transfer coefficients in Nd elastic scattering at 
$E_{lab}^N = 22.7$~MeV. The description of symbols, bands and lines 
 is the same as 
 in Fig.\ref{fig1}.
}
\label{fig3}
\end{figure}

\begin{figure}
\epsfysize=180mm \epsffile{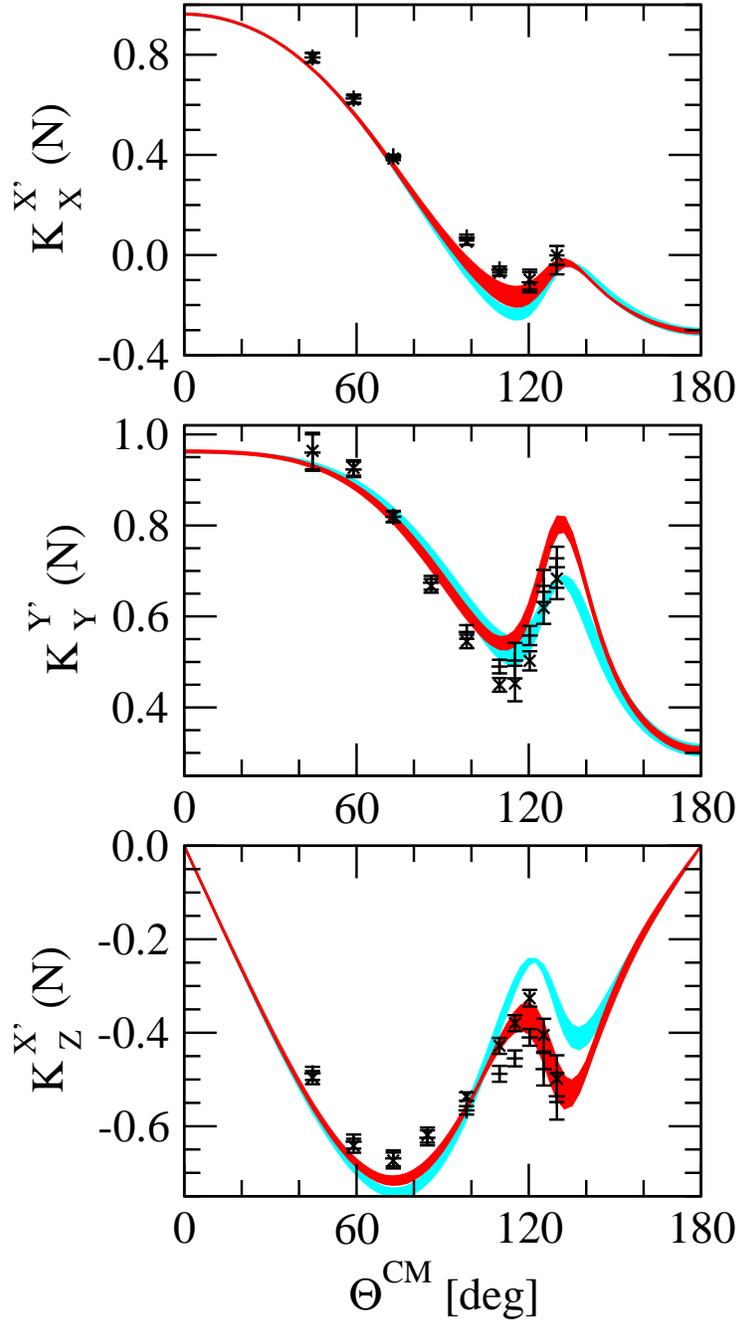}
\caption{
The nucleon to nucleon spin transfer coefficients in Nd elastic scattering at 
$E_{lab}^N = 22.7$~MeV. The description of symbols is the same as 
 in Fig.\ref{fig1}.
The light  band results from theoretical predictions 
obtained with 
the  NLO  chiral potential with different 
cut-off parameters.  The dark band results when in NNLO the 
NN- and 
3N-forces are included.
}
\label{fig4}
\end{figure}

\begin{figure}
\epsfysize=180mm \epsffile{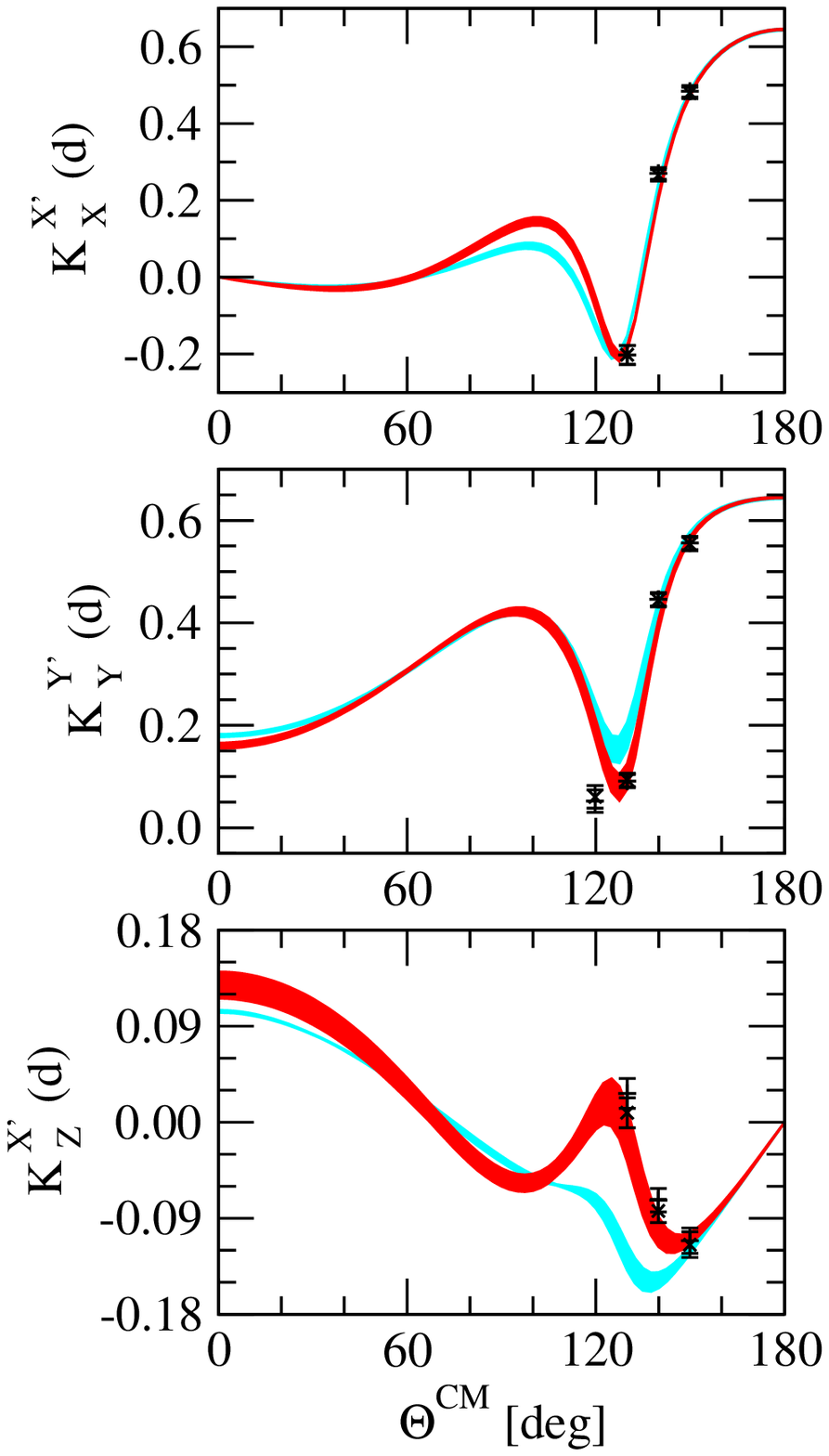}
\caption{
The nucleon to deuteron spin transfer coefficients in Nd elastic scattering at 
$E_{lab}^N = 22.7$~MeV. The description of symbols and bands 
 is the same as 
 in Fig.\ref{fig4}.
}
\label{fig5}
\end{figure}

\begin{figure}
\epsfysize=150mm \epsffile{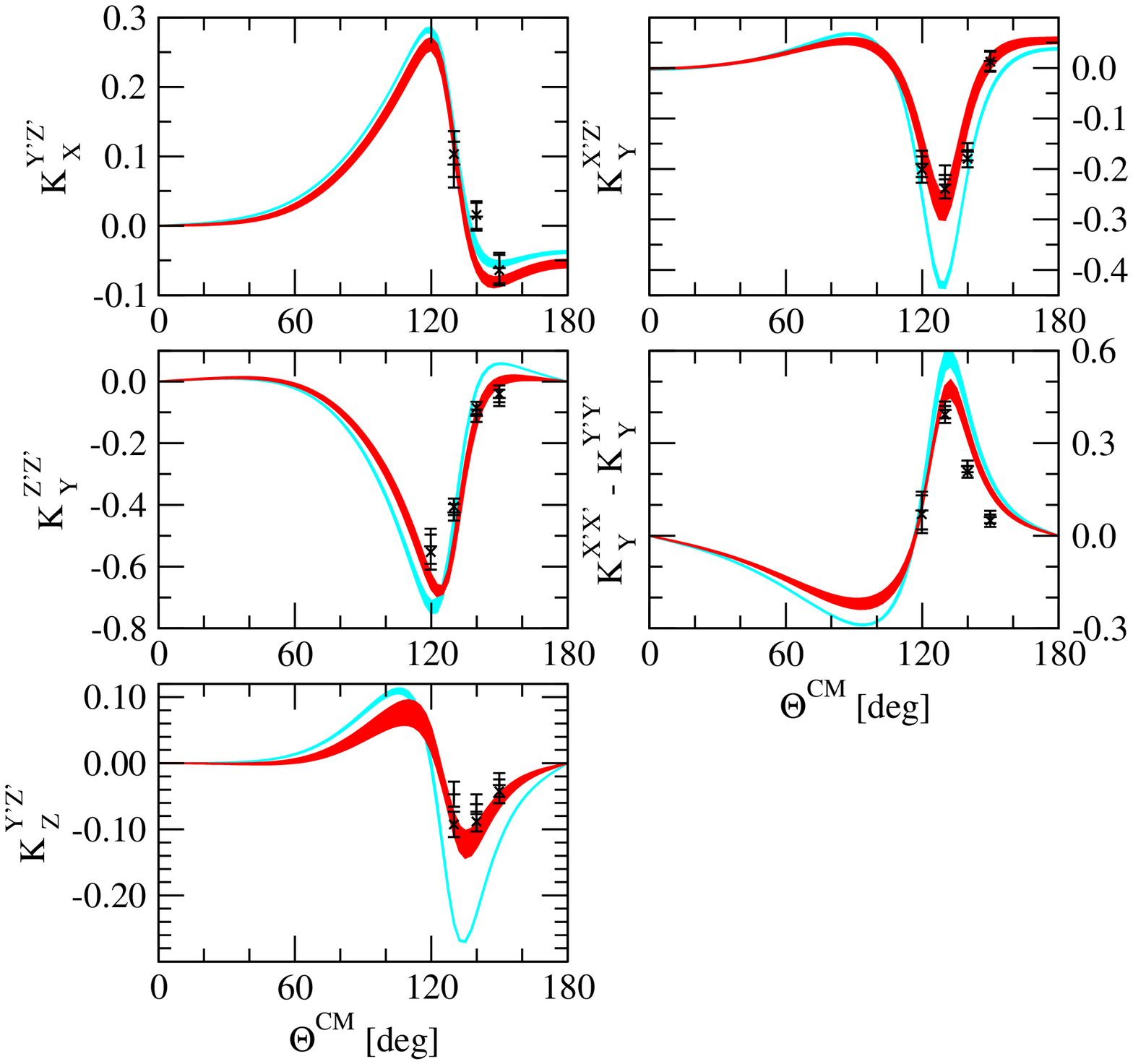}
\caption{
The nucleon to deuteron spin transfer coefficients in Nd elastic scattering at 
$E_{lab}^N = 22.7$~MeV. The description of symbols and bands 
 is the same as 
 in Fig.\ref{fig4}.
}
\label{fig6}
\end{figure}

In Figs.~\ref{fig1}-\ref{fig3} we show our data and compare them to 
theoretical predictions based on (semi)phenomenological 
NN potentials alone or combined with the TM99~\cite{tm99} or 
Urbana IX~\cite{uIX} 3NFs, which have been obtained in 
the Faddeev approach, where we are able to employ also non-local 
 interactions. The 
corresponding comparison for chiral forces is presented in 
Figs.~\ref{fig4}-\ref{fig6}. Comparison with experiment always means 
the Coulomb corrected ``nd'' data.

For the traditional approach based on high quality (semi)phenomenological 
interactions,  we have taken 
the NN potentials AV18, CDBonn, Nijm I and II 
 and combined each of them with the $2\pi$-exchange TM99 3NF,  
adjusting the cut-off parameter of TM99 individually to get 
the experimental triton 
binding energy. The resulting cut-offs for these potentials are respectively 
 5.215, 4.856, 5.120 and 5.072 (in units of the pion mass $m_{\pi}$). From 
the predictions 
of these potentials alone or combined with the TM99 3NF two bands, 
light and dark, respectively, were formed 
and shown in Figs.~\ref{fig1}-\ref{fig3}. 

For the proton to proton spin transfer coefficients (see Fig.~\ref{fig1}) 
the effects of the Coulomb interaction are located in the region of 
c.m. angles around $\theta_{cm} \approx 120^o$. In that region 
the differential cross section has its minimum. 
 They are significant only for 
$K_y^{y'}$ and $K_z^{x'}$ and are practically negligible for 
  $K_x^{x'}$. The realistic potentials alone provide a good description only 
of 
 $K_x^{x'}$  and fail to reproduce the data for $K_y^{y'}$ and $K_z^{x'}$, 
especially around $\theta_{cm} \approx 120^o$. Including the 
TM99 3NF, and 
in case of AV18 also Urbana IX,  changes only slightly the 
predictons for $K_x^{x'}$. For $K_y^{y'}$ and $K_z^{x'}$ the effects of these 
3NFs are significant in the region of c.m. angles 
around $\theta_{cm} \approx 120^o$ and their inclusion leads to a 
good description 
of the data. 

For the corresponding proton to deuteron spin transfer coefficients 
($K_x^{x'}, K_y^{y'}$ and $K_z^{x'}$ - see  Fig.~\ref{fig2})  effects 
of the Coulomb force are practically negligible at angles where data exist 
and they are seen only for  $K_z^{x'}$. For these observables 
 also effects of the TM99 and 
Urbana IX 3NFs are small and the realistic NN potentials alone or 
combined with these 3NFs provide quite good description of the data.

For the proton to tensor-deuteron spin transfer coefficients the Coulomb 
forces are significant for $K_x^{y'z'}$, $K_y^{x'z'}$, and $K_z^{y'z'}$ 
(see Fig.~\ref{fig3}). In case of $K_y^{z'z'}$ and $K_y^{x'x'-y'y'}$ 
(see Fig.~\ref{fig3}) they are negligible at angles where data exist. 
For these two coefficients also 
 the effects of the TM99 and Urbana IX 3NFs are small and 
the NN potentials alone 
provide a good description of data. This is similar  for $K_x^{y'z'}$. 
For $K_y^{x'z'}$ and $K_z^{y'z'}$ the effects of these 3NFs are nonnegligible, 
especially in the region of angles around $\theta_{cm} \approx 120^o$. 
While for $K_y^{x'z'}$  the inclusion of  the TM99 or Urbana IX 3NFs improves 
 the description of the data, in case of $K_z^{y'z'}$ it shifts 
 the theory away from 
 the data points.

Based on the  
 chiral interactions we show in Figs.~\ref{fig4}-\ref{fig6} two bands 
of predictions based on forces derived in NLO and NNLO. Each band is based 
on five predictions obtained with different cut-off combinations: 
(450,500), (600,500), (550,600), (450,700), and (600,700) 
(see \cite{epel03} for more details). 
The results for the chiral NN potentials in NLO  are shown by the light band. 
In NNLO the first time nonzero contributions from chiral three-nucleon 
interactions arise and the predictions based on the full chiral Hamiltonian 
in NNLO are shown by the dark band. 

It is clearly seen that the restriction to NLO only is quite 
insufficient even at low 
energy of our experiment. The predictions based on the 
chiral NN potential obtained 
in this low order are far away from the data. 
 At NNLO one could show the NN force predictions alone. 
We refrain from doing that since this is ambiguous. As is well 
known unitarily transformed 2N forces, which do not affect 
two nucleon observables, lead to different results in the 3N system 
 \cite{polyzu}. It is only the complete 3N Hamiltonian 
which provides unambiguous results for the 3N observables. 
Since the chiral approach systematically improves 
the nuclear force description with increasing order 
and provides strong internal links between 2N forces and forces beyond,  
we deviate here from the usual presentation exhibiting 3N force effects 
separately. This was done above in the standard approach 
since anyhow there is no internal consistency between NN and 3N forces. 
Now a view on Figs.~\ref{fig4}-\ref{fig6} shows that the full NNLO predictions 
lead to a quite 
 good description as the traditional approach including 3N forces. 
The exception are $K_y^{y'}$ for the proton to proton transfer, 
 where the chiral approach differs from the 
data and $K_z^{y'z'}$ for the proton to deuteron transfer, 
 where it leads to an agreement with the data. In the 
standard approach it is opposite. 

It will be of great interest to see the outcome for NNNLO, where the NN forces 
have already been worked out~\cite{nnnlo}. At that order there contribute 
a whole host of parameter free 3NFs including first relativistic 
effects.

\section{Summary and conclusions}

We presented new data for spin transfer coefficients in elastic pd scattering, 
both for the proton to proton and for the  proton to deuteron transfers. 
They have been 
measured using a polarized proton beam with energy $E_p^{lab}=22.7$~MeV. 
The data have been corrected for Coulomb force effects using our theoretical 
framework of a hyperspherical expansion. This leads to ``nd'' data 
to which we compare our theory. In some cases the Coulomb force effects 
are quite 
significant, especially at $\theta_{cm} \approx 120^o$, where the 
differential cross section has its minimum. The theoretical predictions 
are obtained by solving the 3N Faddeev equations with two different 
dynamical inputs. One is the standard approach of so called high 
precision NN forces supplemented by the TM99 and Urbana IX 3NF. The other 
is an 
effective field theory approach constrained by chiral symmetry. In the first 
case where the forces have mostly  phenomenological character we show 
NN force prediction separately in addition to the results obtained by adding 
the 3NFs. 
The inclusion of 3NFs clearly improves the description and the comparison with 
the data is quite successful. In case of the approach 
based on the chiral nuclear forces NLO is quite insufficient, but 
at NNLO the combined dynamics of NN and 3N forces does essentially equally 
well as 
the standard approach. Exceptions are the spin transfer coefficients 
$K_y^{y'}$ from proton to proton and $K_z^{y'z'}$ from proton to deuteron. The 
first is well described 
in the standard approach but not the second one. Just the 
opposite is true in the chiral approach.

\section*{Acknowledgments}
This work has been supported by the  Polish Committee for
Scientific Research under Grant no. 2P03B00825  and by the
U.S. Department of Energy Contract No. DE-AC05-84ER40150 under which the
Southeastern Universities Research Association (SURA) operates the Thomas
Jefferson Accelerator Facility.  
 The numerical
calculations have been performed on the IBM Regatta p690+ of the
NIC in J\"ulich, Germany.


\newpage

\newpage

\newpage

\newpage

\newpage

\newpage

\newpage

\newpage

\newpage

\begin{thebibliography}{99}
\bibitem{AV18} R.B. Wiringa, V.G.J. Stoks, R. Schiavilla,
               Phys. Rev. C{\bf 51}, 38 (1995).

\bibitem{CDBONN} R.~Machleidt, F. Sammarruca, and Y. Song,
                 Phys. Rev. C{\bf 53}, R1483 (1996).

\bibitem{NIJMI} V.G.J. Stoks, R.A.M. Klomp, C.P.F. Terheggen,
                J.J. de Swart, Phys. Rev. C{\bf 49}, 2950 (1994).

\bibitem{wein91} S. Weinberg, Nucl. Phys. {\bf B 363}, 3 (1991).
\bibitem{vankolck94} U. van Kolck, Phys. Rev. C {\bf 49}, 2932 (1994).
\bibitem{epel98} E. Epelbaoum, W. Gl{\"o}ckle, U.-G. Mei{\ss}ner, Nucl. Phys. 
{\bf A 637}, 107 (1998).
\bibitem{epel03} D.R. Entem, R. Machleidt, Phys. Rev. C {\bf 68}, 041001 (2003).

\bibitem{nnnlo} E.\ Epelbaum, W. Gl\"ockle, Ulf-G. Mei{\ss}ner, 
 Nucl. Phys. {\bf{A 747}}, 362 (2005).

\bibitem{epel04a} E. Epelbaum, W. Gl{\"o}ckle, U.-G. Mei{\ss}ner, Eur. Phys. J. 
{\bf A 19}, 401 (2004).

\bibitem{epel04b} E. Epelbaum, W. Gl{\"o}ckle, U.-G. Mei{\ss}ner, 
Eur. Phys. J. {\bf A 19}, 125 (2004).

\bibitem{Friar1993} J.L.~Friar et al., Phys. Lett. B{\bf 311}, 4 (1993).

\bibitem{Nogga1997} A. Nogga, D. H\"uber, H. Kamada, and 
W. Gl\"ockle,  Phys. Lett. B{\bf 409}, 19 (1997).

\bibitem{Argonne} S.C. Pieper, V.R. Pandharipande, R.B. Wiringa, 
and J. Carlson, 
                 Phys. Rev. C{\bf 64}, 014001 (2001).

\bibitem{wit98} H. Wita{\l}a, W. Gl\"ockle, D. H\"uber, J. Golak, 
and H. Kamada, Phys. Rev. Lett. {\bf 81}, 1183 (1998).

\bibitem{glo96} W. Gl\"ockle, H.Wita{\l}a, D.H\"uber, H.Kamada, J.Golak, 
 Phys. Rep. {\bf{274}}, 107 (1996).

\bibitem{nogga03} A. Nogga et al., Phys. Rev. C{\bf 67}, 034004 (2003)

\bibitem{TM} S.A. Coon et al., 
Nucl.  Phys. {\bf A317}, 242 (1979); 
 S.A. Coon and W. Gl\"ockle,  Phys. Rev. C{\bf 23}, 1790 (1981).

\bibitem{uIX} B.S. Pudliner et al., 
  Phys. Rev. C{\bf 56}, 1720 (1997). 

\bibitem{sek02} K. Sekiguchi et al.,  Phys. Rev. C{\bf 65}, 034003 (2002).


\bibitem{wit01} H. Wita{\l}a et al.,  Phys. Rev. C{\bf 63}, 024007 (2001).

\bibitem{abf98} W.P. Abfalterer et al., Phys. Rev. Lett. {\bf 81}, 57 (1998).
\bibitem{wit99} H. Wita{\l}a et al.,  Phys. Rev. C{\bf 59}, 3035 (1999).

\bibitem{hat02} K. Hatanaka et al., Phys. Rev. C{\bf 66}, 044002 (2002).

\bibitem{cad01} R.V. Cadman et al., Phys. Rev. Lett. {\bf 86}, 967 (2001).

\bibitem{KVR01} A. Kievsky, M. Viviani and S. Rosati, Phys. Rev. C {\bf 64},
      024002 (2001)

\bibitem{epel2002} E.\ Epelbaum et al., 
 Phys.\ Rev.\ C {\bf 66}, 064001 (2002).

\bibitem{glomb} A. Glombik et al., AIP Conference Proceedings 334 on Few Body 
Problems in Physics, Williamsburg 1994, ed. F. Gross, (AIP Press, New York, 
1995) p.486.

\bibitem{kretch} W. Kretschmer, private communication.

\bibitem{kretch1} W. Kretschmer et al., AIP Conference Proc. 339 (1995) 335 
(Polarization Phenomena in Nuclear Physics, Bloomington, 1994).

\bibitem{sperisen} F. Sperisen et al., Nucl. Phys.  C{\bf A 422}, 81 (1984).

\bibitem{sydow} L. Sydow  et al., Few-Body Systems {\bf 25}, 133 (1998).

\bibitem{sek04} K. Sekiguchi et al., Phys. Rev. C{\bf 70}, 014001 (2004).

\bibitem{vwitsch} P. Hempen  et al., Phys. Rev. C{\bf 57}, 837 (1998).

%

%
%

\bibitem{wit88}  H.Wita{\l}a, T.Cornelius and W.Gl\"ockle, 
  Few-Body Syst. {\bf{3}}, 123 (1988).

\bibitem{hub97} D. H\"uber, H. Kamada, H. Wita{\l}a, and W. Gl\"ockle, 
 Acta Phys. Pol. B{\bf{28}}, 1677 (1997).

\bibitem{kievsky01}   A. Kievsky, S. Rosati and M. Viviani, Phys. Rev. C
 {\bf 64}, 041001(R)

\bibitem{tm99} S.A. Coon and H.K. Han, 
Few-Body Syst. {\bf 30}, 131 (2001).

\bibitem{polyzu}  W.N. Polyzou, W. Gl\"ockle, 
  Few-Body Syst. {\bf{9}}, 97 (1990).


\end{thebibliography}
\end{document}